# Exploring the Limits of Stability[1]

M. Thoennessen

*National Superconducting Cyclotron Laboratory and*
*Department of Physics & Astronomy, Michigan State University*
*East Lansing, MI 48824, USA*

**Abstract.** The discovery of the currently known most neutron- and proton-rich nuclei is compiled. The compilation includes the year and method of discovery of the last known bound and the first unbound nuclei along the neutron and proton driplines.

**Keywords:** neutron dripline, proton dripline, rare isotopes
**PACS:** 21.10-k

## THE NEUTRON DRIPLINE

The neutron dripline should not be discussed in terms of isotopes as the proton dripline but rather in terms of isotones [1, 2, 3] in order to avoid the odd-even staggering due to pairing. Table 1 shows the lightest stable (with respect to neutron emission) isotones and the isotones beyond the dripline. For the unbound isotones the first observation of the ground-state resonance is listed. Several different experimental methods were used for the discovery of these isotones: spallation (SP), projectile fragmentation (PF), transfer (TR), pion induced (PI) and break-up reactions of radioactive beams (RB). Strictly speaking, the dripline is only known up to N = 9 (or lithium) because no search for $^{13}$Li has been reported.

**TABLE 1.** Last bound (Bou.) and first unbound (Unb.) isotones along the neutron dripline for $3 < N < 12$. The methods of discovery (Me.) are explained in the text.

| N | Bou. | Year | Me. | Author | Ref. | Unb. | Year | Me. | Author | Ref. |
|---|---|---|---|---|---|---|---|---|---|---|
| 4 | $^{6}$He | 1936 | TR | Bjerge | [4] | $^{5}$H | 2001 | RB | Korsheninnikov | [5] |
| 5 | $^{8}$Li | 1935 | TR | Crane | [6] | $^{7}$He | 1967 | TR | Stokes | [7] |
| | | | | | | $^{6}$H | 2003 | PI | Gurov | [8] |
| 6 | $^{8}$He | 1965 | SP | Poskanzer | [9] | $^{7}$H | 2003 | RB | Korsheninnikov | [10] |
| 7 | $^{11}$Be | 1958 | TR | Nurmia | [11] | $^{10}$Li | 1993 | PF | Kryger | [12] |
| | | | | | | $^{9}$He | 2001 | RB | Chen | [13] |
| 8 | $^{11}$Li | 1966 | SP | Poskanzer | [14] | $^{10}$He | 1994 | RB | Korsheninnikov | [15] |
| 9 | $^{14}$B | 1966 | SP | Poskanzer | [14] | $^{13}$Be | 2001 | PF | Thoennessen | [16] |
| 10 | $^{14}$Be | 1973 | SP | Bowman | [17] | — | | | | |
| 11 | $^{17}$C | 1968 | SP | Poskanzer | [18] | $^{16}$B | 2000 | TR | Kalpakchieva | [19] |

**TABLE 2.** Last bound (Bou.) and first unbound (Unb.) isotones along the neutron dripline for 11 < N < 35. The methods of discovery (Me.) are explained in the text.

| N | Bou. | Year | Me. | Author | Ref. | Unb. | Year | Me. | Author | Ref. |
|---|---|---|---|---|---|---|---|---|---|---|
| 12 | $^{17}$B | 1973 | SP | Bowman | [17] | $^{16}$Be | 2003 | PF | Baumann | [20] |
| 13 | $^{19}$C | 1974 | SP | Bowman | [21] | $^{18}$B | 1984 | PF | Musser | [22] |
| 14 | $^{19}$B | 1984 | PF | Musser | [22] | — | | | | |
| 15 | $^{22}$N | 1979 | PF | Westfall | [23] | $^{21}$C | 1985 | PF | Langevin | [24] |
| 16 | $^{22}$C | 1986 | PF | Pougheon | [25] | $^{21}$B | 2003 | PF | Ozawa | [26] |
| 17 | $^{26}$F | 1979 | PF | Westfall | [23] | $^{25}$O | 1985 | PF | Langevin | [24] |
| 18 | $^{27}$F | 1981 | PF | Stevenson | [27] | $^{26}$O | 1990 | PF | Guillemaud-M. | [28] |
| 19 | $^{29}$Ne | 1985 | PF | Langevin | [24] | $^{28}$F | 1999 | PF | Sakurai | [29] |
| 20 | $^{29}$F | 1989 | PF | Guillemaud-M. | [30] | $^{28}$O | 1997 | PF | Tarasov | [31] |
| 21 | $^{31}$Ne | 1996 | PF | Sakurai | [32] | $^{30}$F | 1999 | PF | Sakurai | [29] |
| 22 | $^{31}$F | 1999 | PF | Sakurai | [29] | — | | | | |
| 23 | $^{34}$Na | 1983 | SP | Langevin | [33] | $^{33}$Ne | 2002 | PF | Notani | [34] |
| 24 | $^{34}$Ne | 2002 | PF | Notani | [34] | — | | | | |
| 25 | $^{37}$Mg | 1996 | PF | Sakurai | [32] | $^{36}$Na | 2002 | PF | Notani | [34] |
| 26 | $^{37}$Na | 2002 | PF | Notani | [34] | — | | | | |
| 27 | $^{40}$Al | 1997 | PF | Sakurai | [35] | $^{39}$Mg | 2002 | PF | Notani | [34] |
| 28 | $^{41}$Al | 1997 | PF | Sakurai | [35] | — | | | | |
| 29 | $^{43}$Si | 2002 | PF | Notani | [34] | — | | | | |
| 30 | $^{45}$P | 1990 | PF | Lewitowicz | [36] | — | | | | |
| 31 | $^{46}$P | 1990 | PF | Lewitowicz | [36] | — | | | | |
| 32 | $^{48}$S | 1990 | PF | Lewitowicz | [36] | — | | | | |
| 33 | $^{51}$Ar | 1989 | PF | Guillemaud-M. | [30] | — | | | | |
| 34 | $^{51}$Cl | 1990 | PF | Lewitowicz | [36] | — | | | | |

**TABLE 3.** Last bound (Bou.) and first unbound (Unb.) isotopes along the proton dripline for 3 < Z < 12. The methods of discovery (Me.) are explained in the text.

| Z | Bou. | Year | Me. | Author | Ref. | Unb. | Year | Me. | Author | Ref. |
|---|---|---|---|---|---|---|---|---|---|---|
| 4 | $^{7}$Be | 1938 | TR | Rumbaugh | [37] | $^{6}$Be | 1958 | TR | Bogdanov | [38] |
| 5 | $^{8}$B | 1950 | TR | Alvarez | [39] | $^{7}$B | 1967 | TR | McGrath | [40] |
| 6 | $^{9}$C | 1964 | TR | Cerny | [41] | $^{8}$C | 1974 | TR | Robertson | [42] |
| 7 | $^{12}$N | 1949 | TR | Alvarez | [43] | $^{11}$N | 1974 | TR | Benenson | [44] |
| | | | | | | $^{10}$N | 2002 | TR | Lepine-Szily | [45] |
| 8 | $^{13}$O | 1963 | TR | Barton | [46] | $^{12}$O | 1978 | TR | KeKelis | [47] |
| 9 | $^{17}$F | 1934 | TR | Wertenstein | [48] | $^{16}$F | 1964 | TR | Bryant | [49] |
| | | | | | | $^{15}$F | 1978 | TR | KeKelis | [47] |
| 10 | $^{17}$Ne | 1963 | TR | Barton | [46] | $^{16}$Ne | 1978 | TR | KeKelis | [47] |
| 11 | $^{20}$Na | 1950 | TR | Alvarez | [39] | $^{19}$Na | 1969 | TR | Cerny | [50] |
| | | | | | | $^{18}$Na | 2004 | RB | Zerguerras | [51] |

Table 2 continues along the neutron dripline towards heavier nuclei but the unbound nuclei listed have not been observed. Searches for the existence of these isotones were performed using projectile fragmentation reactions. In this mass region projectile fragmentation (PF) has clearly been the dominant method to explore the neutron dripline. A few isotones were observed by spallation reactions (SP). The listing for the neutron dripline is limited to chlorine or N = 34 because for heavier nuclei the last observed isotones are not approaching the predicted location of the neutron dripline.

It should be noted that the observation of $^{34}$Ne and $^{37}$Na as well as the non-observation of $^{33}$Ne and $^{36}$Na was reported by Lukyanov *et al.* [52] shortly after Notani *et al.* [34]. Notani *et al.* submitted their paper on May 28$^{th}$, 2002 and accepted for publication on July 15$^{th}$, 2002, while the Lukyanov paper was submitted on July 9$^{th}$, 2002.

## THE PROTON DRIPLINE

The proton dripline is much better known than the neutron dripline. Table 3 shows the lightest stable (with respect to protron emission) isotopes and the isotopes beyond the dripline. For the unbound isotopes the first observation of the ground-state resonance is listed. All isotopes were discovered by traditional transfer reactions (TR) with the exception of $^{19}$Na which has been observed by the break-up of a radioactive beam (RB) of $^{20}$Mg [51]. In this mass region, the proton dripline agrees with the line separating bound and unbound nuclei.

$^{15}$F was observed simultaneously by KeKelis *et al.* [47] and Benenson *et al.* [53]. The KeKelis paper was submitted on October 20$^{th}$, 1997 while the Benenson *et al.* paper was submitted on November 28$^{th}$, 1997. They were published next to each other in the same issue of Phys. Rev. C.

Table 4 continues along the proton dripline towards heavier nuclei up to Z = 51. Due to the Coulomb barrier the simple relation between the dripline (as defined as $S_p = 0$) and unbound nuclei does not hold anymore. Nuclei beyond the dripline can have significant lifetimes. The exact location of the dripline is experimentally known for only a few isotopes [2]. The isotopes listed as unbound Table 4 have not been observed in projectile fragmentation reactions and lifetime limits were deduced from the flight-paths through the fragment separator which is typically less than a microsecond. Thus these nuclei are certainly beyond the dripline. The only unbound isotope where the resonance was observed directly by a transfer reaction is $^{39}$Sc [54].

Transfer reactions (TR) for the lighter mass region and projectile fragmentation (PF) for the heavier masses were used to discovers most of the isotopes around the proton dripline. A few nuclei were first observed by fusion evaporation (FE) and spallation (SP) reactions.

Table 5 shows the lightest presently known isotope for $52 < Z < 74$. In this mass region all isotopes along the proton dripline were observed by fusion evaporation and no lifetime limits for any of the isotopes have been determined. Thus the limit of observable nuclei has not been reached [2] although the $S_p = 0$ dripline has clearly been crossed as can be seen from the decay modes (D.) listed.

**TABLE 4.** Last bound (Bou.) and first unbound (Unb.) isotopes along the proton dripline for $11 < Z < 52$. The methods of discovery (Me.) are explained in the text.

| Z | Bou. | Year | Me. | Author | Ref. | Unb. | Year | Me. | Author | Ref. |
|---|---|---|---|---|---|---|---|---|---|---|
| 12 | $^{20}$Mg | 1974 | TR | Robertson | [42] | $^{19}$Mg | 2003 | PF | Frank | [55] |
| 13 | $^{22}$Al | 1982 | TR | Cable | [56] | $^{21}$Al | 1987 | PF | Cerny | [57] |
| 14 | $^{22}$Si | 1987 | PF | Saint-Laurent | [57] | — | | | | |
| 15 | $^{26}$P | 1983 | TR | Cable | [58] | $^{25}$P | 1986 | PF | Langevin | [59] |
| 16 | $^{27}$S | 1986 | TR | Langevin | [59] | — | | | | |
| 17 | $^{31}$Cl | 1977 | TR | Benenson | [60] | $^{30}$Cl | 1986 | PF | Langevin | [59] |
| | | | | | | $^{29}$Cl | 1986 | PF | Langevin | [59] |
| 18 | $^{31}$Ar | 1986 | PF | Langevin | [59] | — | | | | |
| 19 | $^{35}$K | 1976 | TR | Benenson | [61] | $^{34}$K | 1986 | PF | Langevin | [59] |
| | | | | | | $^{33}$K | 1986 | PF | Langevin | [59] |
| 20 | $^{35}$Ca | 1985 | TR | Äystö | [62] | — | | | | |
| 21 | $^{40}$Sc | 1955 | TR | Glass | [63] | $^{39}$Sc | 1988 | TR | Mohar | [54] |
| 22 | $^{39}$Ti | 1990 | PF | Detraz | [64] | $^{38}$Ti | 1996 | PF | Blank | [65] |
| 23 | $^{43}$V | 1987 | PF | Pougheon | [66] | $^{42}$V | 1992 | PF | Borrel | [67] |
| 24 | $^{42}$Cr | 1996 | PF | Blank | [65] | — | | | | |
| 25 | $^{46}$Mn | 1987 | PF | Pougheon | [66] | $^{45}$Mn | 1992 | PF | Borrel | [67] |
| | | | | | | $^{44}$Mn | 1992 | PF | Borrel | [67] |
| 26 | $^{45}$Fe | 1996 | PF | Blank | [65] | — | | | | |
| 27 | $^{50}$Co | 1987 | PF | Pougheon | [66] | $^{49}$Co | 1994 | PF | Blank | [68] |
| 28 | $^{48}$Ni | 2000 | PF | Blank | [69] | — | | | | |
| 29 | $^{55}$Cu | 1987 | PF | Pougheon | [66] | $^{54}$Cu | 1994 | PF | Blank | [68] |
| 30 | $^{54}$Zn | 2005 | PF | Blank | [70] | — | | | | |
| 31 | $^{60}$Ga | 1995 | PF | Blank | [71] | $^{59}$Ga | 2005 | PF | Stolz | [72] |
| 32 | $^{60}$Ge | 2005 | PF | Stolz | [72] | — | | | | |
| 33 | $^{64}$As | 1995 | PF | Blank | [71] | $^{63}$As | 2005 | PF | Stolz | [72] |
| 34 | $^{64}$Se | 2005 | PF | Stolz | [72] | — | | | | |
| 35 | $^{70}$Br | 1978 | FE | Alburger | [73] | $^{69}$Br | 1995 | PF | Blank | [71] |
| | | | | | | $^{68}$Br | 1995 | PF | Blank | [71] |
| 36 | $^{69}$Kr | 1995 | PF | Blank | [71] | — | | | | |
| 37 | $^{74}$Rb | 1977 | SP | D'Auria | [74] | $^{73}$Rb | 1991 | PF | Mohar | [75] |
| | | | | | | $^{72}$Rb | 1995 | PF | Blank | [71] |
| 38 | $^{73}$Sr | 1993 | FE | Batchelder | [76] | — | | | | |
| 39 | $^{76}$Y | 2001 | PF | Kienle | [77] | — | | | | |
| 40 | $^{78}$Zr | 2001 | PF | Kienle | [77] | — | | | | |
| 41 | $^{82}$Nb | 1992 | PF | Yenello | [78] | $^{81}$Nb | 1999 | PF | Janas | [79] |
| 42 | $^{83}$Mo | 1999 | PF | Janas | [79] | — | | | | |
| 43 | $^{86}$Tc | 1992 | PF | Yenello | [78] | $^{85}$Tc | 1999 | PF | Janas | [79] |
| 44 | $^{87}$Ru | 1995 | PF | Rykaczewski | [80] | — | | | | |
| 45 | $^{89}$Rh | 1995 | PF | Rykaczewski | [80] | — | | | | |
| 46 | $^{91}$Pd | 1995 | PF | Rykaczewski | [80] | — | | | | |
| 47 | $^{94}$Ag | 1994 | PF | Hencheck | [81] | — | | | | |
| 48 | $^{97}$Cd | 1978 | SP | Elmroth | [82] | — | | | | |
| 49 | $^{98}$In | 1995 | PF | Rykaczewski | [80] | — | | | | |
| 50 | $^{100}$Sn | 1994 | PF | Schneider | [83] | — | | | | |
| 51 | $^{103}$Sb | 1995 | PF | Rykaczewski | [80] | — | | | | |

**TABLE 5.** Last observed isotopes (Iso.) along the proton dripline for $51 < Z < 73$ and the decay mode (D.) of the first observation.

| Z | Iso. | Year | D. | Author | Ref. | Z | Iso. | Year | D. | Author | Ref. |
|---|---|---|---|---|---|---|---|---|---|---|---|
| 52 | $^{106}$Te | 1981 | $\alpha$ | Schardt | [84] | 73 | $^{155}$Ta | 1999 | p | Uusitalo | [100] |
| 53 | $^{108}$I | 1991 | $\alpha$ | Page | [85] | 74 | $^{158}$W | 1981 | $\alpha$ | Hofmann | [99] |
| 54 | $^{110}$Xe | 1981 | $\alpha$ | Schardt | [84] | 75 | $^{160}$Re | 1992 | p | Page | [101] |
| 55 | $^{112}$Cs | 1994 | p | Page | [86] | 76 | $^{162}$Os | 1989 | $\alpha$ | Hofmann | [102] |
| 56 | $^{114}$Ba | 1997 | $\beta$ | Janas | [87] | 77 | $^{165}$Ir | 1997 | p | Davids | [103] |
| 57 | $^{117}$La | 2001 | p | Soramel | [88] | 78 | $^{166}$Pt | 1996 | $\alpha$ | Bingham | [104] |
| 58 | $^{121}$Ce | 1997 | $\beta$ | Zhankui | [89] | 79 | $^{170}$Au | 2004 | p,$\alpha$ | Kettunen | [105] |
| 59 | $^{121}$Pr | 1990 | $\beta$ | Bogdanov | [90] | 80 | $^{171}$Hg | 2004 | $\alpha$ | Kettunen | [105] |
| 60 | $^{125}$Nd | 1999 | $\beta$ | Xu | [91] | 81 | $^{176}$Tl | 2004 | p | Kettunen | [105] |
| 61 | $^{128}$Pm | 1999 | $\beta$ | Xu | [91] | 82 | $^{180}$Pb | 1996 | $\alpha$ | Toth | [106] |
| 62 | $^{129}$Sm | 1999 | $\beta$ | Xu | [91] | 83 | $^{184}$Bi | 2003 | $\alpha$ | Andreyev | [107] |
| 63 | $^{130}$Eu | 2004 | p | Davids | [92] | 84 | $^{188}$Po | 1999 | $\alpha$ | Andreyev | [108] |
| 64 | $^{134}$Gd | 2004 | - | Woods | [93] | 85 | $^{191}$At | 2003 | $\alpha$ | Kettunen | [109] |
| 65 | $^{135}$Tb | 2004 | p | Woods | [93] | 86 | $^{195}$Rn | 2001 | $\alpha$ | Kettunen | [110] |
| 66 | $^{139}$Dy | 1999 | $\beta$ | Xu | [91] | 87 | $^{199}$Fr | 1999 | $\alpha$ | Tagaya | [111] |
| 67 | $^{140}$Ho | 1999 | p | Rykaczewski | [94] | 88 | $^{201}$Ra | 2005 | $\alpha$ | Uusitalo | [112] |
| 68 | $^{144}$Er | 2003 | - | Karny | [95] | 89 | $^{206}$Ac | 1998 | $\alpha$ | Eskola | [113] |
| 69 | $^{145}$Tm | 1998 | p | Batchelder | [96] | 90 | $^{209}$Th | 1996 | $\alpha$ | Ikezoe | [114] |
| 70 | $^{149}$Yb | 2001 | $\beta$ | Xu | [97] | 91 | $^{212}$Pa | 1997 | $\alpha$ | Mitsuoka | [115] |
| 71 | $^{150}$Lu | 1993 | p | Sellin | [98] | 92 | $^{217}$U | 2000 | $\alpha$ | Malyshev | [116] |
| 72 | $^{154}$Hf | 1981 | $\alpha$ | Hofmann | [99] | | | | | | |

## REMARKS

The present compilation contains only refereed publications. Publications in conference proceedings are not included. Sometimes this separation is not obvious. For example, $^{164}$Ir has been reported in Eur. Phys. J. A. [117] in 2002, and although apparently a contribution to a conference it was not identified as such. This nucleus has not been reported in a regular refereed publication since, so it was omitted from the present compilation and $^{165}$Ir first observed in 1997 is still the lightest iridium isotope [103].

A compilation based on a literature search as the present one is probably not perfect. Corrections and feedback are very welcome and should be communicated to the author (thoennessen@nscl.msu.edu). The most recent entry included is dated from June $17^{th}$, 2005 [70].

## ACKNOWLEDGMENTS

This work was supported by the National Science Foundation grant number PHY01-10253.

# REFERENCES

1. P. G. Hansen and J. A. Tostevin, *Ann. Rev. Nucl. Part. Sci.*, **53**, 219 (2003).
2. M. Thoennessen, *Rep. Prog. Phys.* **67**, 1187 (2004).
3. M. Thoennessen, *Eur. Phys. J. A direct*, DOI:10.1140/epjad/i2005–06–001–9 (2005).
4. T. Bjerge, *Nature* **147**, 865 (1936).
5. A. A. Korsheninnikov *et al.*, *Phys. Rev. Lett.* **87**, 092501 (2001).
6. H. R. Crane *et al.*, *Phys. Rev.* **47**, 971 (1935).
7. R. H. Stokes and P. G. Young, *Phys. Rev. Lett.* **18**, 611 (1967).
8. Y. B. Gurov *et al.*, *JETP Lett.* **78**, 183 (2003).
9. A. Poskanzer *et al.*, *Phys. Rev. Lett.* **15**, 1030 (1965).
10. A. A. Korsheninnikov *et al.*, *Phys. Rev. Lett.* **90**, 082501 (2003).
11. M. J. Nurmia and R. W. Fink, *Phys. Rev. Lett.* **1**, 23 (1958).
12. R. A. Kryger *et al.*, *Phys. Rev. C* **47**, R2439 (1993).
13. L. Chen *et al.*, *Phys. Lett. B* **505**, 21 (2001).
14. A. Poskanzer *et al.*, *Phys. Rev. Lett.* **17**, 1271 (1966).
15. A. A. Korsheninnikov *et al.*, *Phys. Lett. B* **326**, 31 (1994).
16. M. Thoennessen *et al.*, *Phys. Rev. C* **63**, 014308 (2001).
17. J. D. Bowman *et al.*, *Phys. Rev. Lett.* **31**, 614 (1973).
18. A. Poskanzer *et al.*, *Phys. Lett.* **27B**, 414 (1968).
19. R. Kalpakchieva *et al.*, *Eur. Phys. J. A* **7**, 451 (2000).
20. T. Baumann *et al.*, *Phys. Rev. C* **67**, 061303(R) (2003).
21. J. D. Bowman *et al.*, *Phys. Rev. C* **9**, 836 (1974).
22. J. A. Musser and J. D. Stevenson, *Phys. Rev. Lett.* **53**, 2544 (1984).
23. G. D. Westfall *et al.*, *Phys. Rev. Lett.* **43**, 1859 (1979).
24. M. Langevin *et al.*, *Phys. Lett.* **150B**, 71 (1985).
25. F. Pougheon *et al.*, *Europhys. Lett.* **2**, 505 (1986).
26. A. Ozawa *et al.*, *Phys. Rev. C* **67**, 014610 (2003).
27. J. D. Stevenson and P. B. Price, *Phys. Rev. C* **24**, 2102 (1981).
28. D. Guillemaud-Mueller *et al.*, *Phys. Rev. C* **41**, 937 (1990).
29. H. Sakurai *et al.*, *Phys. Lett. B* **448**, 180 (1999).
30. D. Guillemaud-Mueller *et al.*, *Z. Phys. A* **332**, 189 (1989).
31. O. Tarasov *et al.*, *Phys. Lett. B* **409**, 64 (1997).
32. H. Sakurai *et al.*, *Phys. Rev. C* **54**, R2802 (1996).
33. M. Langevin *et al.*, *Phys. Lett.* **125B**, 116 (1983).
34. M. Notani *et al.*, *Phys. Lett. B* **542**, 49 (2002).
35. H. Sakurai *et al.*, *Nucl. Phys. A* **616**, 311c (1997).
36. M. Lewitowicz *et al.*, *Z. Phys. A* **335**, 117 (1990).
37. L. H. Rumbaugh *et al.*, *Phys. Rev.* **54**, 657 (1938).
38. G. F. Bogdanov *et al.*, *J. Nuclear Energy II* **8**, 148 (1958).
39. L. W. Alvarez *et al.*, *Phys. Rev.* **80**, 519 (1950).
40. R. L. McGrath *et al.*, *Phys. Rev. Lett.* **19**, 1442 (1967).
41. J. Cerny *et al.*, *Phys. Rev. Lett.* **13**, 726 (1964).
42. R. G. H. Robertson *et al.*, *Phys. Rev. Lett.* **32**, 1207 (1974).
43. L. W. Alvarez *et al.*, *Phys. Rev.* **75**, 1815 (1949).
44. W. Benenson *et al.*, *Phys. Rev. C* **9**, 2130 (1974).
45. A. Lepine-Szily *et al.*, *Phys. Rev. C* **65**, 054318 (2002).
46. R. Barton *et al.*, *Can. J. Phys.* **41**, 12 (1963).
47. G. J. KeKelis *et al.*, *Phys. Rev. C* **17**, 1929 (1978).
48. L. Wertenstein *et al.*, *Nature* **133**, 564 (1934).
49. H. C. Bryant *et al.*, *Nucl. Phys.* **53**, 97 (1964).
50. J. Cerny *et al.*, *Phys. Rev. Lett.* **22**, 612 (1969).
51. T. Zerguerras *et al.*, *Eur. Phys. J. A* **20**, 389 (2004).
52. S. Lukyanov *et al.*, *J. Phys. G* **28**, L41 (2002).
53. W. Benenson *et al.*, *Phys. Rev. C* **17**, 1939 (1978).
54. M. Mohar *et al.*, *Phys. Rev. C* **38**, 737 (1988).

55. N. Frank *et al.*, *Phys. Rev. C* **68**, 054309 (2003).
56. M. D. Cable *et al.*, *Phys. Rev. C* **26**, 1778 (1982).
57. M. G. Saint-Laurent *et al.*, *Phys. Rev. Lett.* **59**, 33 (1987).
58. M. D. Cable *et al.*, *Phys. Lett.* **123B**, 25 (1983).
59. M. Langevin *et al.*, *Nucl. Phys. A* **455**, 149 (1986).
60. W. Benenson *et al.*, *Phys. Rev. C* **15**, 1187 (1977).
61. W. Benenson *et al.*, *Phys. Rev. C* **13**, 1479 (1976).
62. J. Aysto *et al.*, *Phys. Rev. Lett.* **55**, 1384 (1985).
63. N. W. Glass and J. R. Richardson, *Phys. Rev.* **98**, 1251 (1955).
64. C. Detraz *et al.*, *Nucl Phys. A* **519**, 529 (1990).
65. B. Blank *et al.*, *Phys. Rev. Lett.* **77**, 2893 (1996).
66. F. Pougheon *et al.*, *Z. Phys. A* **327**, 17 (1987).
67. V. Borrel *et al.*, *Z. Phys. A* **344**, 135 (1992).
68. B. Blank *et al.*, *Phys. Rev. C* **50**, 2398 (1994).
69. B. Blank *et al.*, *Phys. Rev. Lett.* **84**, 1116 (2000).
70. B. Blank *et al.*, *Phys. Rev. Lett.* **94**, 232501 (2005).
71. B. Blank *et al.*, *Phys. Rev. Lett.* **74**, 4611 (1995).
72. A. Stolz *et al.*, *Eur. Phys. J. A direct*, DOI:10.1140/epjad/i2005–06–192–y, accept. for publ. Phys. Lett. B (2005).
73. D. E. Alburger, *Phys. Rev. C* **18**, 1875 (1978).
74. J. D'Auria *et al.*, *Phys. Lett.* **66B**, 233 (1977).
75. M. F. Mohar *et al.*, *Phys. Rev. Lett.* **66**, 1571 (1991).
76. J. C. Batchelder *et al.*, *Phys. Rev. C* **48**, 2593 (1993).
77. P. Kienle *et al.*, *Prog. Part. Nucl. Phys.* **46**, 73 (2001).
78. S. Yennello *et al.*, *Phys. Rev. C* **46**, 2620 (1992).
79. Z. Janas *et al.*, *Phys. Rev. Lett.* **82**, 295 (1999).
80. K. Rykaczewski *et al.*, *Phys. Rev. C* **52**, R2310 (1995).
81. M. Hencheck *et al.*, *Phys. Rev. C* **50**, 2219 (1994).
82. T. Elmroth *et al.*, *Nucl. Phys. A* **304**, 493 (1978).
83. R. Schneider *et al.*, *Z. Phys. A* **348**, 241 (1994).
84. D. Schardt *et al.*, *Nucl. Phys. A* **368**, 153 (1981).
85. R. Page *et al.*, *Z. Phys. A* **338**, 295 (1991).
86. R. Page *et al.*, *Phys. Rev. Lett.* **72**, 1798 (1994).
87. Z. Janas *et al.*, *Nucl Phys. A* **627**, 119 (1997).
88. F. Soramel *et al.*, *Phys. Rev. C* **63**, 031304 (2001).
89. L. Zhankui *et al.*, *Phys. Rev. C* **56**, 1157 (1997).
90. D. D. Bogdanov *et al.*, *Sov. J. Nucl. Phys.* **52**, 229 (1990).
91. S-W. Xu *et al.*, *Phys. Rev. C* **60**, 061302 (1999).
92. C. N. Davids *et al.*, *Phys. Rev. C* **69**, 011302(R) (2004).
93. P. J. Woods *et al.*, *Phys. Rev. C* **69**, 051302 (2004).
94. K. Rykaczewski *et al.*, *Phys. Rev. C* **60**, 011301 (1999).
95. M. Karny *et al.*, *Phys. Rev. Lett.* **90**, 012502 (2003).
96. J. C. Batchelder *et al.*, *Phys. Rev. C* **57**, R1042 (1998).
97. S. Xu *et al.*, *Eur. Phys. J. A* **12**, 1 (2001).
98. P. Sellin *et al.*, *Phys. Rev. C* **47**, 1933 (1993).
99. S. Hofmann *et al.*, *Z. Phys. A* **299**, 281 (1981).
100. J. Uusitalo *et al.*, *Phys. Rev. C* **59**, R2975 (1999).
101. R. Page *et al.*, *Phys. Rev. Lett.* **68**, 1287 (1992).
102. S. Hofmann *et al.*, *Z. Phys. A* **333**, 107 (1989).
103. C. N. Davids *et al.*, *Phys. Rev. C* **55**, 2255 (1997).
104. C. R. Bingham *et al.*, *Phys. Rev. C* **54**, R20 (1996).
105. H. Kettunen *et al.*, *Phys. Rev. C* **69**, 054323 (2004).
106. K. Toth *et al.*, *Z. Phys. A* **355**, 225 (1996).
107. A. N. Andreyev *et al.*, *Eur. Phys. J. A* **18**, 55 (2003).
108. A. N. Andreyev *et al.*, *Eur. Phys. J. A* **6**, 381 (1999).
109. H. Kettunen *et al.*, *Eur. Phys. J. A* **17**, 537 (2003).

110. H. Kettunen *et al.*, *Phys. Rev. C* **63**, 044315 (2001).
111. Y. Tagaya *et al.*, *Eur. Phys. J. A* **5**, 123 (1999).
112. J. Uusitalo *et al.*, *Phys. Rev. C* **71**, 024306 (2005).
113. K. Eskola *et al.*, *Phys. Rev. C* **57**, 417 (1998).
114. H. Ikezoe *et al.*, *Phys. Rev. C* **54**, 2043 (1996).
115. S. Mitsuoka *et al.*, *Phys. Rev. C* **55**, 1555 (1997).
116. O. Malyshev *et al.*, *Eur. Phys. J. A* **8**, 295 (2000).
117. H. Mahmud *et al.*, *Eur. Phys. J. A* **15**, 85 (2002).